\documentclass[useAMS,usenatbib,twcolumn,usegraphicx]{mn2e}
\usepackage{times}

\newcommand{\beq}{\begin{equation}}
\newcommand{\eeq}{\end{equation}}
\newcommand{\beqar}{\begin{eqnarray}}
\newcommand{\eeqar}{\end{eqnarray}}

\title{Alfv\'{e}n QPOs in Magnetars}
\author[H.~Sotani, K.~D. Kokkotas, \& N.~Stergioulas]
{H.~Sotani$^1$\thanks{E-mail:sotani@astro.auth.gr},
K.~D.~Kokkotas$^{1,2}$\thanks{E-mail:kokkotas@auth.gr},
and N.~Stergioulas$^1$\thanks{E-mail:niksterg@astro.auth.gr}
\\
  $^1$Department of Physics, Aristotle University of Thessaloniki,
  Thessaloniki 54124, Greece \\
  $^2$Theoretical Astrophysics, University of T\"{u}bingen, Auf der Morgenstelle 10,
  72076, T\"{u}bingen, Germany}
\begin{document}

\maketitle

\label{firstpage}

%%%%%%%%%%%%%%%%%%%%%%%%%%%%%%%%%%%%%%%%%%%%%%%%%%%%%%%%%%%%%%%%%%%%%%%%%%%%
% Abstract
\begin{abstract}
  We investigate torsional Alfv\'{e}n oscillations of relativistic stars with a 
  global dipole magnetic field, via two-dimensional numerical simulations. We find 
  that a) there exist two families of 			
  quasi-periodic oscillations (QPOs) with harmonics at integer multiples of the fundamental 
  frequency, b) the lower-frequency QPO is related to the region of closed field 
  lines, near the equator, while the higher-frequency QPO is generated near the 
  magnetic axis, c) the QPOs are long-lived, d) for the chosen form of dipolar 
  magnetic field, the frequency ratio of the lower to upper fundamental QPOs is $\sim 
  0.6$, independent of the equilibrium model or of the strength of the magnetic
  field, and e) within a representative sample of equations of state and of 
  various magnetar masses, the Alfv\'{e}n QPO frequencies are given by
  accurate empirical relations that depend only on the compactness of the 
  star and on the magnetic field strength. The lower and upper QPOs can be 
  interpreted as corresponding to the edges or turning points of an Alfv\'{e}n
  continuum, according to the model proposed by Levin (2007). Several of the 
  low-frequency QPOs observed in the X-ray tail of SGR 1806-20 can readily be 		
  identified with 
  the Alfv\'{e}n QPOs we compute. In particular, one could identify the 18Hz and
  30Hz observed frequencies with the fundamental lower and upper QPOs, 
  correspondingly, while the observed frequencies of 
  92Hz and 150Hz are then integer multiples of the fundamental upper QPO 
  frequency (three times and five times, correspondingly). With this 
  identification, we obtain an upper limit on the strength of magnetic field 
  of SGR 1806-20 (if is dominated by a dipolar component)  
  between $\sim3$ and $7\times 10^{15}$G. Furthermore, we show that 
  an identification of the observed frequency of 26Hz with the frequency of the 
  fundamental torsional $\ell=2$ oscillation of the magnetar's crust 
  is compatible with a magnetar mass of about $1.4$ to 1.6$M_\odot$ and an EOS that is 	very stiff (if the magnetic field strength is near its upper limit) or 		
  moderately stiff (for lower values of the	magnetic field). 
\end{abstract}
%%%%%%%%%%%%%%%%%%%%%%%%%%%%%%%%%%%%%%%%%%%%%%%%%%%%%%%%%%%%%%%%%%%%%%%%%%%%


%%%%%%%%%%%%%%%%%%%%%%%%%%%%%%%%%%%%%%%%%%%%%%%%%%%%%%%%%%%%%%%%%%%%%%%%%%%%
%%%%%%%%%%%%%%%%%%%%%%%%%%%%%%%%%%%%%%%%%%%%%%%%%%%%%%%%%%%%%%%%%%%%%%%%%%%%
\section{Introduction}
\label{sec:Intro}
%%%%%%%%%%%%%%%%%%%%%%%%%%%%%%%%%%%%%%%%%%%%%%%%%%%%%%%%%%%%%%%%%%%%%%%%%%%%
%%%%%%%%%%%%%%%%%%%%%%%%%%%%%%%%%%%%%%%%%%%%%%%%%%%%%%%%%%%%%%%%%%%%%%%%%%%%

The phenomenon of Soft Gamma Repeaters (SGRs) may allow us in the near future
to determine fundamental properties of strongly magnetized, compact stars.
Already, there exist at least two sources in which
quasi-periodic oscillations (QPOs) have been observed in their X-ray tail,
following the initial discovery by \cite{Israel2005}, see \cite{WS2006} for a recent review.
The frequency of many of these oscillations is similar to what one would expect 
for torsional modes of the solid crust of a compact star. This observation
is in support of the proposal that SGRs are magnetars (compact objects with
very strong magnetic fields) \citep{DT1992}. During an SGR event,
torsional oscillations in the solid crust of the star could be excited \citep{Duncan1998},
leading to the observed frequencies in the X-ray tail. However, not all of
the observed frequencies fit the above picture. For example, the three lowest
observed frequencies for SGR 1806-20 are 18, 26, 30Hz. Only one of these could
be the fundamental, $\ell=2,m=0$ torsional frequency of the crust, as the 
first overtone has a much higher frequency. \cite{Levin2006} stressed the
importance of crust-core coupling by a global magnetic field and of the
existence of an Alfv\'en continuum, while \citet{GSA2006} considered model with simplified geometry, in which Alfv\'{e}n oscillations form a discrete spectrum
of normal modes, that could be associated with the observed low-frequency QPOS. 
In  \cite{Levin2007}, the existence of a continuum was stressed further 
and it was shown that the edges or turning points of the continuum can yield long-lived QPOs. In addition, 
numerical simulations showed that drifting QPOs within the continuum become
amplified near the frequencies of the crustal normal modes. Within this
model, Levin suggested a likely identification of the 18Hz QPO in SGR 1806-20 
with the lowest frequency of the MHD continuum or its first overtone.
The above results were obtained in toy models with simplified geometry and Newtonian 
gravity.

In this Letter, we perform two-dimensional numerical simulations of linearized 
Alfv\'{e}n oscillations in magnetars. Our model improves on the previously 
considered toy models in various ways: relativistic gravity is assumed, various 
realistic equations of state (EOS) are considered and a consistent dipolar magnetic
field is constructed. We do not consider the presence of a solid crust, but
only examine the response of the ideal magnetofluid to a chosen initial perturbation.

Spherical stars have generally two type of oscillations, {\it spheroidal}
with polar parity and {\it toroidal} with axial parity.
The observed QPOs in SGR X-ray tails may originate from toroidal oscillations,
since these could be excited more easily than poloidal
oscillations, because they do not involve density variations.
In Newtonian theory, there have been several investigations of torsional
oscillations in the crust region of neutron stars (see e.g.,
\cite{Lee2007} for reference). On the other hand, only few studies 
have taken general relativity into account \citep{Messios2001,Sotani2007a,Sotani2007b,SA2007,Vavoulidis2007}.

SGRs produce giant flares with peak luminosities of $10^{44}$ --
$10^{46}$ erg/s, which display a decaying tail for several hundred
seconds.  Up to now, three giant flares have been detected, SGR 0526-66
in 1979, SGR 1900+14 in 1998, and SGR 1806-20 in 2004.  The timing
analysis of the latter two events revealed several QPOs
in the decaying tail, whose frequencies are approximately
18, 26, 30, 92, 150, 625, and 1840 Hz for SGR 1806-20,
and 28, 53, 84, and 155 Hz for SGR 1900+14, see \cite{WS2006}.

In \cite{Sotani2007a} (hereafter Paper~I), it was suggested that some of the observational data of SGRs
could agree with the crustal torsional oscillations,
if, e.g., frequencies lower than 155 Hz are identified with the fundamental
oscillations of different harmonic index $\ell$, while higher frequencies
are identified with overtones. However, in Paper~I and above, 
it will be quite challenging to identify all observed QPO frequencies
with only crustal torsional oscillations.  For example, it is difficult to
explain all of the frequencies of 18, 26 and 30 Hz for SGR 1806-20 with
crustal models, because the actual spacing of torsional oscillations
of the crust is larger than the difference between these two
frequencies. Similarly, the spacing between the 625Hz and a possible 720Hz QPO
in SGR 1806-20 may be too small to be explained by consecutive
overtones of crustal torsional oscillations. 

One can notice, however, that the frequencies of 30, 92 and 150 Hz 
in SGR 1806-20 are in near {\it integer ratios}. As we will show below, the numerical results presented
in this Letter are compatible with this observation, as we
find two families of QPOs (corresponding to the edges or turning points of
a continuum) with harmonics at near integer multiples. Furthermore, our 
results are compatible with the ratio of 0.6 between the 18 and 30Hz frequencies,
if these are identified, as we suggest, with the edges (or turning points)
of the Alfv\'{e}n continuum. With this identification,
we can set an upper limit to the dipole magnetic field of $\sim3$ to $~7\times 10^{15}$G. 
If the drifting QPOs of the
continuum are amplified at the fundamental frequency of the crust, and
the latter is assumed to be the observed 26Hz for SGR 1806-20, then 
our results  are compatible with a magnetar mass of about $1.4$ to 1.6$M_\odot$ and an EOS that is very stiff (if the magnetic field strength
is near its upper limit) or moderately stiff (for lower values of the
magnetic field). 

Unless otherwise noted, we adopt units of $c=G=1$,
where $c$ and $G$ denote the speed of light and the gravitational
constant, respectively, while the metric signature is $(-,+,+,+)$.

%%%%%%%%%%%%%%%%%%%%%%%%%%%%%%%%%%%%%%%%%%%%%%%%%%%%%%%%%%%%%%%%%%%%%%%%%%%%
%%%%%%%%%%%%%%%%%%%%%%%%%%%%%%%%%%%%%%%%%%%%%%%%%%%%%%%%%%%%%%%%%%%%%%%%%%%%
\section[]{Numerical Setup}
\label{sec:II}
%%%%%%%%%%%%%%%%%%%%%%%%%%%%%%%%%%%%%%%%%%%%%%%%%%%%%%%%%%%%%%%%%%%%%%%%%%%%
%%%%%%%%%%%%%%%%%%%%%%%%%%%%%%%%%%%%%%%%%%%%%%%%%%%%%%%%%%%%%%%%%%%%%%%%%%%%

The general-relativistic equilibrium stellar model is assumed to be spherically symmetric and static,
i.e. a solution of the well-known TOV equations for and
perfect fluid and metric described the line element
\begin{equation}
 ds^2 = -e^{2\Phi(r)}dt^2 + e^{2\Lambda(r)}dr^2 + r^2(d\theta^2 + \sin^2\theta d\phi^2).
\end{equation}
We neglect the influence of the magnetic field on the structure of the star,
since the magnetic field energy, ${\cal E_M}$, is orders of magnitudes smaller than the
gravitational binding energy, ${\cal E_G}$, for magnetic field strengths 
considered realistic for magnetars, ${\cal E_M}/{\cal E_G}\approx 10^{-4}(B/(10^{16} {\rm G}))^2$.
For simplicity, we assume that the magnetic field is a pure dipole (toroidal 
magnetic fields will be treated elsewhere, see \cite{Sotani2007c}. Details on
the numerical method for constructing the magnetic field, as well as representative figure of
the magnetic field lines, can be found in Paper~I. 

MHD oscillations of the above equilibrium model are described by the 
linearized equations of motion and the magnetic induction equations, presented
in detail in Papers~I and II (we neglect perturbations in the spacetime metric, as
these couple weakly to toroidal modes in a spherically symmetric background).
The perturbative equations presented in Papers~I and II can readily be 
converted from an eigenvalue problem to the form of a two-dimensional time-evolution problem,
by defining a displacement ${\cal Y}(t,r,\theta)$ due to the toroidal motion
(the coefficient of shear viscosity is set to $\mu=0$, 
as we neglect the presence of a solid crust in the present work).
The contravariant {\it coordinate component} of the perturbed four-velocity, $\delta u^{\phi}$, is
then related to time derivative of 
${\cal Y}$ through
\begin{equation}
 \delta u^{\phi} = e^{-\Phi}\partial_t {\cal Y}(t,r,\theta).
\end{equation}
The two-dimensional evolution equation for ${\cal Y}(t,r,\theta)$ is
\begin{eqnarray}
 {\cal A}_{tt}\frac{\partial^2 {\cal Y}}{\partial t^2}
     &=& {\cal A}_{20}\frac{\partial^2 {\cal Y}}{\partial r^2}
     + {\cal A}_{11} \frac{\partial^2 {\cal Y}}{\partial r \partial \theta}
     + {\cal A}_{02} \frac{\partial^2 {\cal Y}}{\partial \theta^2} \nonumber \\
     &+& {\cal A}_{10} \frac{\partial {\cal Y}}{\partial r}
     + {\cal A}_{01} \frac{\partial {\cal Y}}{\partial \theta}
     + \varepsilon_D {\cal D}_4 {\cal Y},
\label{eq2D}
\end{eqnarray}
where ${\cal A}_{tt}$, ${\cal A}_{20}$, ${\cal A}_{11}$, ${\cal A}_{02}$,
${\cal A}_{10}$, and ${\cal A}_{01}$ are functions of $r$ and $\theta$,
given by
\begin{eqnarray}
 {\cal A}_{tt} &=& \left[\epsilon + p + \frac{{a_1}^2}{\pi r^4} \cos^2\theta 
     + \frac{{{a_1}'}^2}{4\pi r^2} e^{-2\Lambda}\sin^2\theta \right] e^{-2(\Phi - \Lambda)}, \\
 {\cal A}_{20} &=& \frac{{a_1}^2}{\pi r^4}\cos^2\theta , \\
 {\cal A}_{11} &=& -\frac{a_1 {a_1}'}{\pi r^4}\sin\theta \cos\theta, \\
 {\cal A}_{02} &=& \frac{{{a_1}'}^2}{4\pi r^4}\sin^2\theta, \\
 {\cal A}_{10} &=& \left(\Phi' - \Lambda' \right) \frac{{a_1}^2}{\pi r^4} \cos^2\theta
     + \frac{a_1 {a_1}'}{2\pi r^4} \sin^2\theta, \\
 {\cal A}_{01} &=& \left[\frac{a_1}{\pi r^4}\left(2\pi j_1 - \frac{a_1}{r^2}\right)e^{2\Lambda}
     + \frac{3{{a_1}'}^2}{4\pi r^4} \right]\sin\theta\cos\theta,
\end{eqnarray}
and where $a_1(r)$ and $j_1(r)$ are the radial components of the electromagnetic four-potential
and the four-current, respectively. In the above equations, a 
prime denotes a partial derivative with respect to the radial coordinate. 

In our numerical scheme, we employ the 2nd-order, iterative Crank-Nicholson 
scheme. A numerical instability, which sets in after many oscillations, was
treated by adding a 4th-order Kreiss-Oliger dissipation term \citep{KO1973},
shown as $\varepsilon_D {\cal D}_4 {\cal Y}$ in Equation (\ref{eq2D}) above. We
experimented with various values of the dissipation coefficient $\varepsilon_D$
and found the evolution to be stable for values as small as a few times $10^{-5}$.
We verified that in this limit, the solution becomes independent of the
strength of the numerical dissipation, as the ${\cal D}_4$ Kreiss-Oliger
dissipation operator introduces an error of higher order than the 2nd-order iterative Crank-Nicholson scheme.
The numerical grid we use is equidistant,
covering only the interior of the star, with (typically) 50 radial zones
and 40 angular zones (we also compared our results to simulations with
100x80 points). The boundary conditions are: (a) ${\cal Y}=0$ at $r=0$ (regularity), (b) ${\cal Y}_{,r}=0$ at $r=R$ (vanishing traction) (c) ${\cal Y}_{,\theta}=0$ at $\theta=0$ (axisymmetry) and (d) ${\cal Y}=0$ at $\theta=\pi/2$ (equatorial plane symmetry of the $\ell=2$ initial data). We obtain the same results if you use a grid that extends to $\theta=\pi$. We have also evolved initial data with $\ell=3$, for which the appropriate boundary condition at $\theta=\pi/2$ is ${\cal Y}_{,\theta}=0$.

%%%%%%%%%%%%%%%%%%%%%%%%%%%%%%%%%%%%%%%%%%%%%%%%%%%%%%%%%%%%%%%%%%%%%%%%%%%%
%%%%%%%%%%%%%%%%%%%%%%%%%%%%%%%%%%%%%%%%%%%%%%%%%%%%%%%%%%%%%%%%%%%%%%%%%%%%
\section[]{Alfv\'{e}n QPOs}
\label{sec:III}
%%%%%%%%%%%%%%%%%%%%%%%%%%%%%%%%%%%%%%%%%%%%%%%%%%%%%%%%%%%%%%%%%%%%%%%%%%%%
%%%%%%%%%%%%%%%%%%%%%%%%%%%%%%%%%%%%%%%%%%%%%%%%%%%%%%%%%%%%%%%%%%%%%%%%%%%%

\begin{figure}
\begin{center}
\includegraphics[width=55mm]{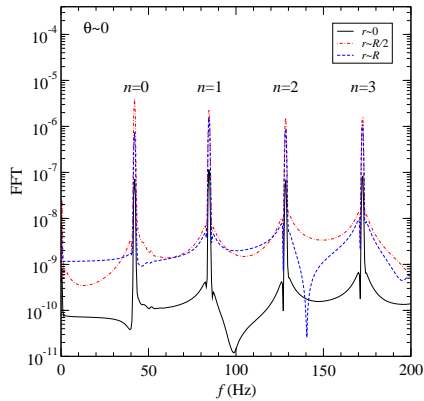}
\end{center}
%\vspace{5mm}
 \caption{FFT of the MHD oscillations at $\theta\sim 0$ (magnetic
 axis).
 Three lines in each figure correspond to different radial positions
 $r\sim 0$, $R/2$, and $R$). A fundamental ($n=0$) QPO and
 several overtones (nearly integer multiples) are clearly present.}
  \label{Fig:FFT}
\end{figure}

\begin{figure}
\begin{center}
\includegraphics[width=55mm]{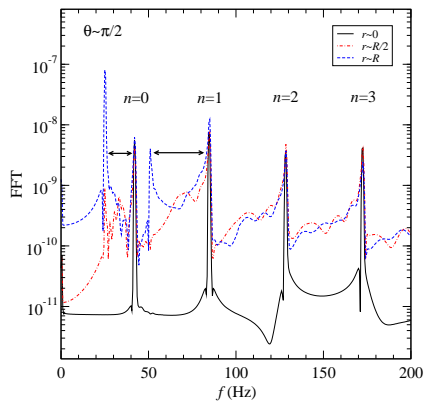}
\end{center}
%\vspace{5mm}
 \caption{Same as Fig. \ref{Fig:FFT}, but at  $\theta\sim \pi/2$
 (magnetic equator). A second family of QPOs is present. Arrows
 indicate several continuous parts, of which only the first is distinct
 from the others, which partially overlap.}
  \label{Fig:FFT2}
\end{figure}

\begin{figure}
\begin{center}
\begin{tabular}{cc}
\includegraphics[width=36mm]{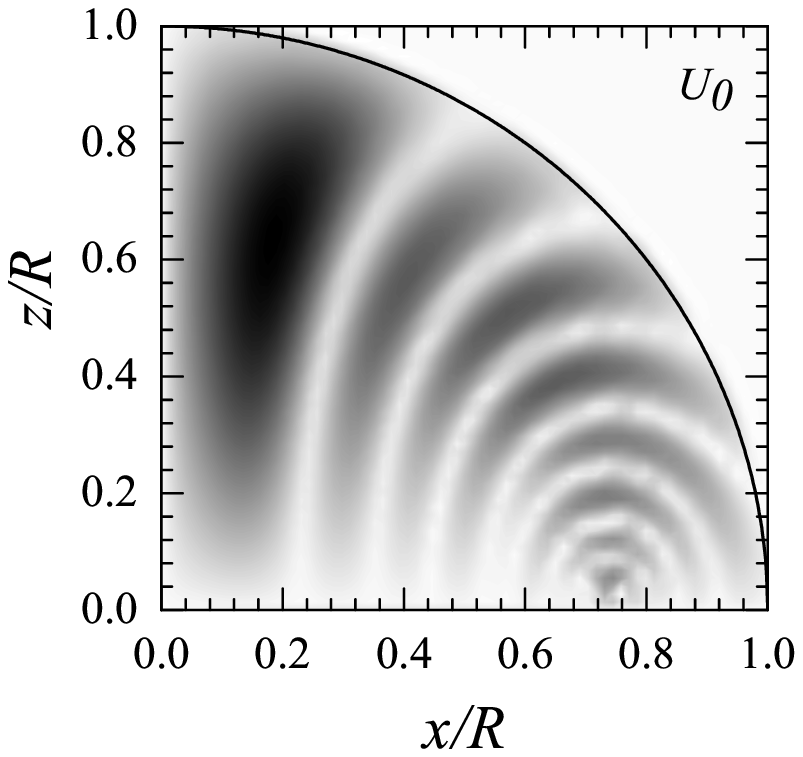} &
\includegraphics[width=36mm]{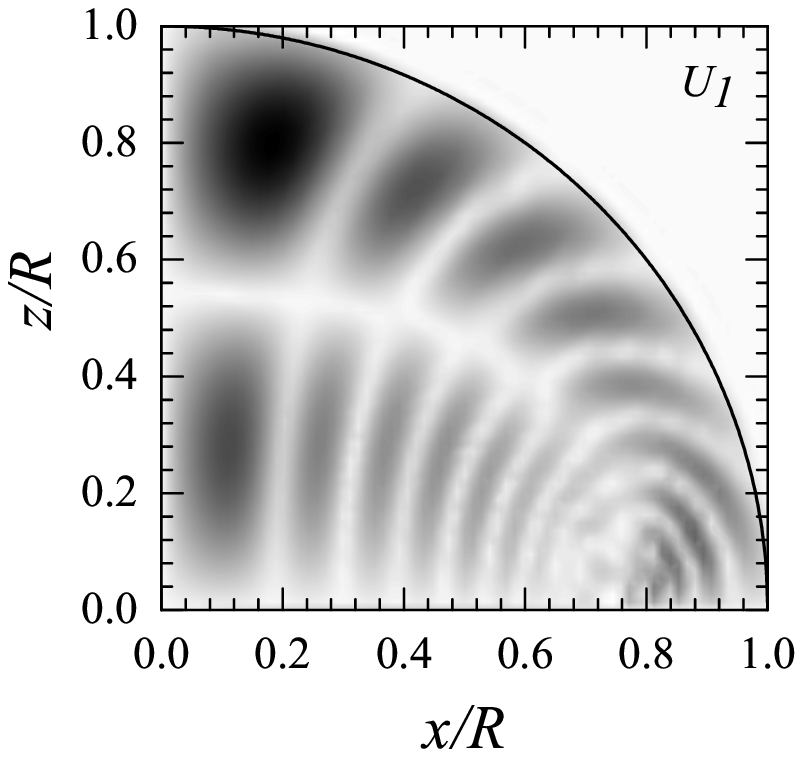} \\
\includegraphics[width=36mm]{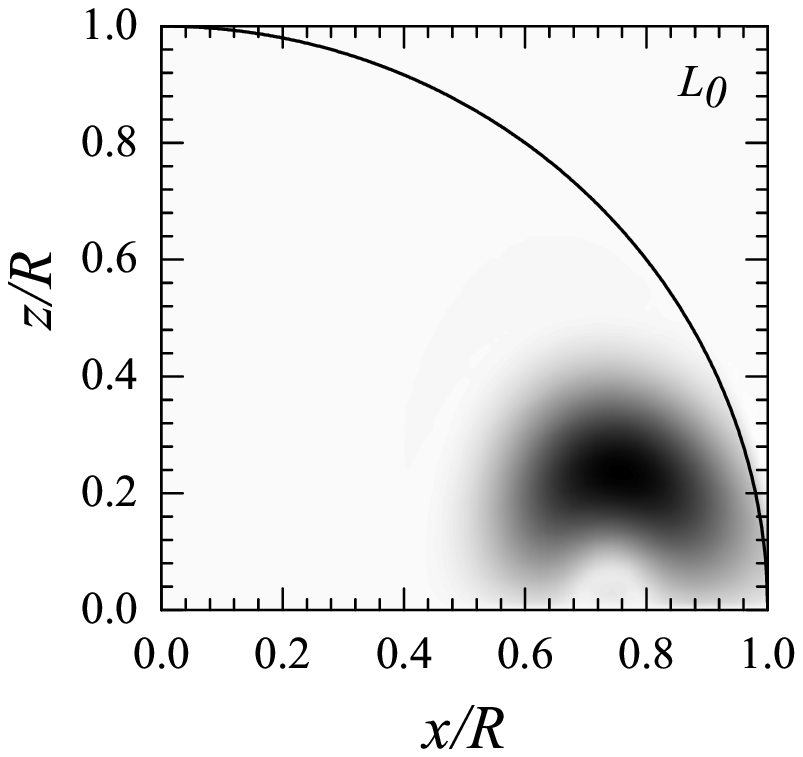} &
\includegraphics[width=36mm]{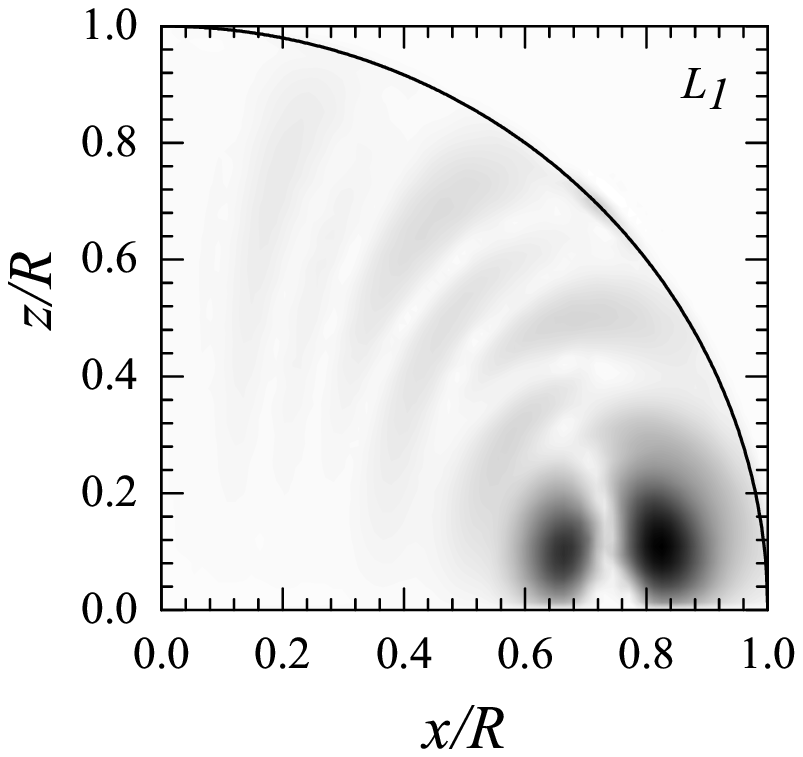} \\
\end{tabular}
\end{center}
%\vspace{5mm}
 \caption{Distribution of effective amplitude of several Alfv\'{e}n QPOs
 (see text for details). The grayscale map varies from white (zero amplitude)
 to black (maximum amplitude).}
  \label{Fig:Effective}
\end{figure}

As a fist equilibrium model, we examined a polytropic stellar model with $\Gamma=2$,
whose mass and radius are $M=1.4M_{\odot}$ and $R=14.16$ km, respectively,
with a dipole magnetic field strength of $B\equiv B_\mu = 4\times 10^{15}$ G,
where $B_{\mu}$ is a typical strength. As initial data,
we use the numerical eigenfunction of the $\ell=2,m=0$ fundamental mode,
for the truncated system presented in Paper~II. The grid size is $100 \times 80$ and
$\varepsilon_D=10^{-3}$, while the total simulation time is 2s. By computing the FFT of $\partial_t{\cal Y}$
(which is proportional to $\delta u^\phi$) at various points inside the
star, we obtain the following results: Examining the FFT at three different
radial locations ($r\sim 0, R/2, R$) at $\theta \sim 0$ (magnetic axis), Fig. \ref{Fig:FFT}, we 
observe a number of narrow frequency peaks. The strength of the peaks
is several orders of magnitude larger than the FFT continuum. The overtones
($n>0$) are nearly {\it integer multiples} of the fundamental frequency
($n=0)$. The corresponding FFTs at the same radial points, but at
$\theta \sim \pi/2$ (magnetic equator) are shown in Fig. \ref{Fig:FFT2}, where, in addition to 
the family of frequency peaks observed in Fig. \ref{Fig:FFT}, one can also 
observe a second family of frequency peaks, mainly at $r\sim R$, for
which the fundamental frequency is at a ratio of 0.6 with respect to the
fundamental frequency in  Fig. \ref{Fig:FFT}. The fundamental
frequency of this family, as well as the first overtone (again an integer
multiple) are clearly visible, while the other overtones seem to be buried
inside a continuum of other frequencies.

Using Levin's toy model for the Alfv\'{e}n continuum in magnetars, the 
above numerical results can be interpreted as follows: The fundamental
frequency peaks of the two families are QPOs generated by the edges or turning
points of an  Alfv\'{e}n continuum (henceforth we call these two 
frequencies as the fundamental lower and upper QPOs and denote them
as $L_0$ and $U_0$). We denote the overtones (which are nearly integer
multiples) as $L_n$ and $U_n$, where $n>0$. For the chosen form of the
magnetic field, the extent of the first continuum, between the fundamental
$L_0$ and $U_0$ QPOs is at lower frequencies than the continuous 
frequency intervals corresponding to the overtones, which all overlap partially.

Because of the Alfv\'{e}n continuum, the phase of the oscillations at the
QPO frequencies is not constant, but varies throughout the star. Since the
magnetic field is axisymmetric, the axis is both a turning point and edge 
of the continuum. Near the magnetic axis, the background magnetic field varies slowly,
which allows for the upper QPOs to remain nearly coherent for
a large number of oscillations. Since $\delta u^\phi$ is only a coordinate
component, we also obtained FFTs of $r \sin\theta \partial_t {\cal Y}$ (which
is proportional to the physical velocity in a unit basis) at all grid points.
In two-dimensional simulations of fluid modes in non-magnetized stars, it
has been shown that the amplitude of the FFT of physical variables at 
every point in the numerical grid, at a chosen normal mode frequency,
is correlated to the shape of the eigenfunction of a given mode
\citep{Stergioulas2004}.
Similarly, we use the magnitude of the FFT of $r \sin\theta \partial_t {\cal Y} $
at every point in the grid to define an {\it effective amplitude} for 
each QPO (which, of course, will be time-varying, due to the absorption of 
individual modes in the continuum). In Fig. \ref{Fig:Effective}, we show
the effective amplitude of the fundamental upper QPO, $U_0$ and its first
overtone $U_1$ as well as for the corresponding lower 
QPOs $L_0$ and $L_1$, obtained after a simulation with a $50\times 40$ grid, 
$\varepsilon_D=2\times 10^{-5}$ and a duration of 2.1s. 

The effective amplitude
for $U_0$ has a maximum near the magnetic axis, while there are also 
{\it nodal lines}
along certain magnetic field lines. For the overtone, $U_1$, there exists an
additional nodal line, starting perpendicular near the {\it magnetic axis}, which
divides the region of maximum amplitude near the magnetic axis roughly
in half. Similarly (not shown here), each successive overtone corresponds to
an additional ``horizontal'' nodal line, dividing the region of maximum amplitude
near the magnetic axis into roughly equidistant parts. This agrees well 
with the fact that the frequencies of the overtones are nearly integer 
multiples of the fundamental frequency. 

The effective amplitude of the fundamental lower QPO, $L_0$, is practically
limited to a region that is only somewhat larger than the region of 
{\it closed magnetic field lines} inside the star. For the first 
overtone $L_1$, a nodal line divides the region of maximum amplitude into two parts.

\begin{figure}
\begin{center}
\includegraphics[width=70mm]{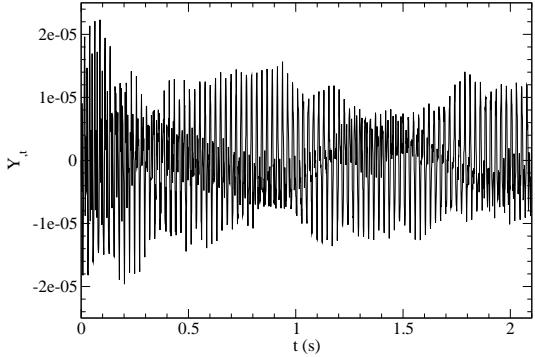} 
\end{center}
%\vspace{5mm}
 \caption{Time evolution of $\partial_t {\cal Y}$ at the location inside the star
 where the effective amplitude of the fundamental upper QPO,
$U_0$ attains its maximum value. The amplitude of the 
oscillations barely diminish with time}
  \label{Fig:evol1}
\end{figure}

In Fig. \ref{Fig:evol1} we show the evolution of $\partial_t {\cal Y}$
at the location inside the star where the effective amplitude of the fundamental upper QPO,
$U_0$ attains its maximum value. It is evident that the QPO is long-lived,
since the amplitude of the oscillations barely diminishes with time.

%%%
\begin{table*}
 \centering
% \begin{minipage}{70mm}
  \caption{Frequencies of lower and upper Alfv\'{e}n QPOs and their ratios,
  for a representative sample of equilibrium models, constructed with various
  EOSs and masses and for a magnetic field strength of $B=B_\mu$ (see
  text for details).}
\label{Tab:eos1}
  \begin{tabular}{@{}lcccccccc@{}}
  \hline
   Model  & M/R  & $f_{L_0}$ (Hz) & $f_{U_0}$ (Hz) & ratio 
                 & $f_{L_1}$ (Hz) & $f_{U_1}$ (Hz) & ratio &  $f_{U_2}$ (Hz) \\
 \hline
 A+DH$_{14}$    & 0.218	&  15.4  &  25.0  &  0.616  &  30.7  &  49.4  &  0.621 & 74.4\\
 A+DH$_{16}$    & 0.264 &  11.7  &  18.3  &  0.639  &  23.5  &  35.7  &  0.658 & 54.0\\
 WFF3+DH$_{14}$ & 0.191	&  17.9  &  29.8  &  0.601  &  36.2  &  59.2  &  0.611 & 89.8\\
 WFF3+DH$_{18}$ & 0.265 &  11.7  &  18.0  &  0.650  &  23.5  &  35.5  &  0.662 & 53.3\\
 APR+DH$_{14}$  & 0.171 &  20.4  &  34.1  &  0.598  &  41.3  &  68.6  &  0.602 & 104.6\\
 APR+DH$_{20}$  & 0.248 &  12.8  &  20.6  &  0.621  &  26.0  &  40.3  &  0.645 & 61.0\\
 L+DH$_{14}$    & 0.141 &  23.7  &  40.8  &  0.581  &  47.5  &  81.6  &  0.582 & 123.8\\
 L+DH$_{20}$    & 0.199 &  16.4  &  27.8  &  0.590  &  33.1  &  54.7  &  0.605 & 82.6\\
\hline
\end{tabular}
%\end{minipage}
\end{table*}
%%%

\begin{figure}
\begin{center}
\includegraphics[width=55mm]{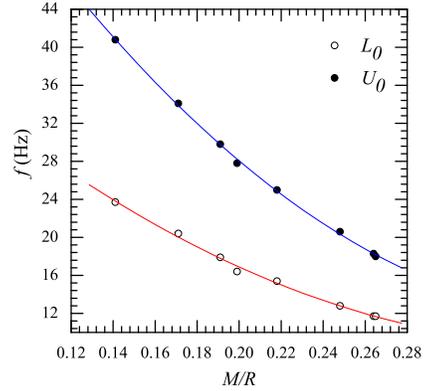} 
\end{center}
%\vspace{5mm}
 \caption{Quadratic fits in terms of the compactness of the star, $M/R$, 
 of the lower and upper fundamental Alfv\'{e}n QPO frequencies, obtained
 for a representative sample of equilibrium models with various EOSs and
 masses. The magnetic field was set to $B=B_\mu$. }
  \label{Fig:fit}
\end{figure}

%%%%%%%%%%%%%%%%%%%%%%%%%%%%%%%%%%%%%%%%%%%%%%%%%%%%%%%%%%%%%%%%%%%%%%%%%%%%
%%%%%%%%%%%%%%%%%%%%%%%%%%%%%%%%%%%%%%%%%%%%%%%%%%%%%%%%%%%%%%%%%%%%%%%%%%%%
\section{Realistic EOS and Empirical Relations}
\label{sec:IV}
%%%%%%%%%%%%%%%%%%%%%%%%%%%%%%%%%%%%%%%%%%%%%%%%%%%%%%%%%%%%%%%%%%%%%%%%%%%%
%%%%%%%%%%%%%%%%%%%%%%%%%%%%%%%%%%%%%%%%%%%%%%%%%%%%%%%%%%%%%%%%%%%%%%%%%%%%

We have obtained the lower and upper Alfv\'{e}n QPO frequencies for a
representative sample of magnetar models with realistic (tabulated) EOSs
and various masses. These equilibrium models and their detailed 
properties have already been presented in Paper~I. In Table \ref{Tab:eos1}
we summarize our numerical results for $B=B_\mu$, for a somewhat smaller 
sample of models, than the one considered in Paper~I. As in the case of the polytropic model of
Sec. \ref{sec:III}, the overtones are nearly integer multiples
of the fundamental frequency
\begin{eqnarray}
 f_{L_n} &\simeq& (n+1)f_{L_0}, \\
 f_{U_n} &\simeq& (n+1)f_{U_0}, 
\end{eqnarray}
with an accuracy on the order of 1\% or better. In addition, the ratio 
of the lower to upper QPO frequencies (also shown in Table \ref{Tab:eos1}) 
roughly agrees with the value of 0.6 for the polytropic model. A quadratic fit 
in terms of $M/R$ gives:
\begin{equation}
\frac{f_{L_0}}{f_{U_0}}\simeq0.62\left [ 1-1.08 \left(\frac{M}{R}\right)+4.52\left(\frac{M}{R}\right)^2\right ],
\end{equation}
with an accuracy of better than $2\%$. Similarly, a quadratic fit for the
ratio of the lower to upper first overtone frequencies yields
\begin{equation}
\frac{f_{L_1}}{f_{U_1}}\simeq0.58\left [ 1-0.54 \left(\frac{M}{R}\right)+3.88\left(\frac{M}{R}\right)^2\right ],
\end{equation}
with a similar accuracy. 

We find that the frequencies of the Alfv\'{e}n QPOs scale {\it linearly} 
with the strength of the magnetic field (with an accuracy on the order of 
1\% or better) at least for magnetic field strengths on the order of $B_\mu$. 

Within the representative sample of the equilibrium models we display in
Table \ref{Tab:eos1}, the frequencies of all lower and upper Alfv\'{e}n QPOs 
(with $n=0,1,2, ...$) are given by the following two empirical relations
\begin{eqnarray}
 f_{L_n} ({\rm Hz})&\simeq& 48.9(n+1)\left[1 - 4.51 \left(\frac{M}{R}\right) +6.18 \left(\frac{M}{R}\right)^2\right] \nonumber \\ 
 &&\times \left(\frac{B}{4\times 10^{15} {\rm G}} \right) ,
     \label{eq:empirical-low} \\
 f_{U_n} ({\rm Hz})&\simeq& 86.1(n+1)\left[1 - 4.58 \left(\frac{M}{R}\right) +6.06 \left(\frac{M}{R}\right)^2\right] \nonumber \\
       &&\times \left(\frac{B}{4\times 10^{15} {\rm G}} \right) ,
     \label{eq:empirical-up}
\end{eqnarray}
with an accuracy of less than about 4\%. The quadratic fits in terms of 
$M/R$, for $B=B_\mu$ is shown in Fig. \ref{Fig:fit}. We emphasize that
in all of the above relations, the ratio $M/R$ is dimensionless, in gravitational
units ($c=G=1$). To restore units, it has to be replaced by $GM/Rc^2$.

%%%%%%%%%%%%%%%%%%%%%%%%%%%%%%%%%%%%%%%%%%%%%%%%%%%%%%%%%%%%%%%%%%%%%%%%%%%%
%%%%%%%%%%%%%%%%%%%%%%%%%%%%%%%%%%%%%%%%%%%%%%%%%%%%%%%%%%%%%%%%%%%%%%%%%%%%
\section{CONSTRAINTS ON MAGNETIC FIELD STRENGTH AND EOS}
\label{sec:V}
%%%%%%%%%%%%%%%%%%%%%%%%%%%%%%%%%%%%%%%%%%%%%%%%%%%%%%%%%%%%%%%%%%%%%%%%%%%%
%%%%%%%%%%%%%%%%%%%%%%%%%%%%%%%%%%%%%%%%%%%%%%%%%%%%%%%%%%%%%%%%%%%%%%%%%%%%

Several of the low-frequency QPOs observed in the X-ray tail of SGR 1806-20 can readily be identified with the Alfv\'{e}n QPOs we compute. In particular, one could identify the 18Hz and 30Hz observed frequencies
with the fundamental lower and upper QPOs, correspondingly,
while the observed frequencies of 92Hz and 150Hz would then be 
integer multiples of the fundamental upper QPO frequency
(three times and five times, correspondingly). With this identification, our empirical relations  Eqs. (\ref{eq:empirical-low}) and (\ref{eq:empirical-up}) constrain the magnetic field strength of SGR 1806-20 (if is dominated by a dipolar component) to be 
between $~3\times 10^{15}$G and $~7\times 10^{15}$G.
Furthermore, an identification of the observed frequency of 26Hz with the frequency of the fundamental torsional $\ell=2$ oscillation of the magnetar's crust (Eq. (79) of Paper~I) implies a very stiff equation of state and a mass
of about 1.4 to 1.6$M_\odot$.
For example, for the $1.4M_{\odot}$ model constructed with EOS L+DH, one obtains  the following frequencies: $_2t_0=25.8$Hz, $f_{L_0}=17.5$Hz, $f_{U_0}=30.0$Hz, $f_{U_3}=90.1$Hz and $f_{U_5}=150.2$Hz, for $B=2.94\times 10^{15}$ G.

Alternatively, one could also identify the 18Hz and
30Hz observed frequencies with overtones (which are also at a near 0.6 ratio).
In this case, the strength of the magnetic field derived above is only an
upper limit and the actual magnetic field may be weaker. Then, if one assumes
that the observed frequency of 26Hz is due to the fundamental $\ell=2$ crust
mode for a weak magnetic field, our numerical data agree best with a 
~1.4$M_\odot$ model constructed with an EOS of moderate stiffness. For 
example, for the 1.4$M_\odot$ model constructed with the APR+DH EOS one obtains $_2t_0=25.9$Hz, $f_{L_1}=17.7$Hz, $f_{U_1}=30.0$Hz,
$f_{U_{5}}=90.1$Hz and $f_{U_{9}}=150.1$Hz, for $B=1.77\times 10^{15}$ G.

%%%%%%%%%%%%%%%%%%%%%%%%%%%%%%%%%%%%%%%%%%%%%%%%%%%%%%%%%%%%%%%%%%%%%%%%%%%%
%%%%%%%%%%%%%%%%%%%%%%%%%%%%%%%%%%%%%%%%%%%%%%%%%%%%%%%%%%%%%%%%%%%%%%%%%%%%
\section{Discussion}
\label{sec:V}
%%%%%%%%%%%%%%%%%%%%%%%%%%%%%%%%%%%%%%%%%%%%%%%%%%%%%%%%%%%%%%%%%%%%%%%%%%%%
%%%%%%%%%%%%%%%%%%%%%%%%%%%%%%%%%%%%%%%%%%%%%%%%%%%%%%%%%%%%%%%%%%%%%%%%%%%%

We have already verified that our main QPO frequencies agree with frequencies obtained with an independent, fully nonlinear numerical code \citep{CerdaDuran2007}, for the same initial model. We caution, however, 
that we have not yet considered the crust-core interaction, different magnetic 
field topologies or the coupling to the exterior magnetosphere. These 
effects have to be taken into account and already \cite{Sotani2007c}, find that the observed QPOs could lead to constraints on the magnetic field topology. To complete the picture, a three-dimensional numerical simulation, that 
includes a proper coupling of the crust to the MHD interior and to the exterior 
magnetosphere will be required and our current results provide a good 
starting point. Extensive details of our computations will be 
presented in \cite{Sotani2007d}.

%\newpage
%%%%%%%%%%%%%%%%%%%%%%%%%%%%%%%%%%%%%%%%%%%%%%%%%%%%%%%%%%%%%%%%%%%%%%
\section*{Acknowledgments}
%%%%%%%%%%%%%%%%%%%%%%%%%%%%%%%%%%%%%%%%%%%%%%%%%%%%%%%%%%%%%%%%%%%%%%

It is a pleasure to thank Yuri Levin for stressing the importance of
the MHD continuum. We are grateful to Pablo C\'erda-Duran and Toni Font
for comparing our main frequency results with their fully nonlinear code
and to Demosthenes Kazanas for helpful discussions. We also thank the
participants of the NSDN Magnetic Oscillations Workshop in Tuebingen 
for useful interactions. This work was supported by the Marie-Curie grant
MIF1-CT-2005-021979, the Pythagoras II program of the Greek Ministry
of Education and Religious Affairs, the EU network ILIAS and the
German Foundation for Research (DFG) via the SFB/TR7 grant.

%%%%%%%%%%%%%%%%%%%%%%%%%%%%%%%%%%%%%%%%%%%%%%%%%%%%%%%%%%%%%%%%%%%%%%%%%%%%
%%%%%%%%%%%%%%%%%%%%%%%%%%%%%%%%%%%%%%%%%%%%%%%%%%%%%%%%%%%%%%%%%%%%%%%%%%%%
%\begin{thebibliography}{999}


\begin{thebibliography}{99}
%%%%%%%%%%%%%%%%%%%%%%%%%%%%%%%%%%%%%%%%%%%%%%%%%%%%%%%%%%%%%%%%%%%%%%%%%%%%
%%%%%%%%%%%%%%%%%%%%%%%%%%%%%%%%%%%%%%%%%%%%%%%%%%%%%%%%%%%%%%%%%%%%%%%%%%%%

%\bibitem[\protect\citeauthoryear{Carroll et al.}{1986}]{Carroll1986}
%   Carroll B.W., Zweibel E.G., Hansen C., McDermot P.N., Savedoff M.P.,
%   Thomas J.H., Van Horn H.M., 1986, ApJ, 305, 767
%   Carroll B.W., et. al., 1986, ApJ, 305, 767

%\bibitem[\protect\citeauthoryear{Douchin \& Haensel}{2001}]{DH2001}
%   Douchin F., Haensel P.,  2001, A\& A, 380, 151

\bibitem[\protect\citeauthoryear{Cerd\'a-Dur\'an et al.}{2007}]{CerdaDuran2007}
    Cerd\'a-Dur\'an P., Sotani H., Stergioulas N., Font J.A., 2007, in preparation
    
\bibitem[\protect\citeauthoryear{Duncan \& Thompson}{1992}]{DT1992}
    Duncan R.C., Thompson C., 1992, ApJ, 392, L9

\bibitem[\protect\citeauthoryear{Duncan}{1998}]{Duncan1998}
    Duncan R.C.,1998, ApJ, 1998,  498, L45

\bibitem[\protect\citeauthoryear{Glampedakis et al.}{2006}]{GSA2006}
    Glampedakis K., Samuelsson L., Andersson N., 2006, MNRAS, 371, L74

%\bibitem[\protect\citeauthoryear{Hansen \& Cioffi}{1980}]{HC1980}
%   Hansen C., Cioffi D.F, 1980, ApJ, 238, 740

%\bibitem[\protect\citeauthoryear{Hurley et al.}{1999}]{SGR1900}
%    Hurley K., et al.,
    %Cline T., Mazets E., Barthelmy S., Butterworth P.,
    %Marshall F., Palmer D., Aptekar R., Golenetskii S., Il'Inskii V.,
    %Frederiks D., McTiernan J., Gold R., Trombka J.,
%    1999, Nature, 397, L41

\bibitem[\protect\citeauthoryear{Israel et al.}{2005}]{Israel2005}
    Israel G. et al., 2005, ApJ, 628, L53

%\bibitem[\protect\citeauthoryear{Kermer et al.}{1986}]{KLR1986}
%    Kerner W., Lerbinger K., Riedel K., 1986, Phys. Fluids 29, 2975

%\bibitem[\protect\citeauthoryear{Kouveliotou et al.}{1998}]{SGR1806}
%    Kouveliotou C., et al.,
    %Dieters S., Strohmayer T., van Paradijis J.,
    %Fishman G.J., Meegan C.A., Hurley K., Kommers J., Smith I.,
    %Frail D., Murakami T.,
%    1998, Nature, 393, L235


\bibitem[\protect\citeauthoryear{Kreiss \& Oliger}{1973}]{KO1973}
    Kreiss H.-O., Oliger J., 1973, Methods for the approximate solution of time dependent problems,
    GARP Publications Series 10

\bibitem[\protect\citeauthoryear{Lee}{2007}]{Lee2007}
    Lee U., 2007, MNRAS, 374, 1015

%\bibitem[\protect\citeauthoryear{Leins}{1994}]{Leins1994}
%    Leins M., 1994, PhD Thesis, University of T\"ubingen

\bibitem[\protect\citeauthoryear{Levin}{2006}]{Levin2006}
    Levin Y., 2006, MNRAS, 368, L35
    
\bibitem[\protect\citeauthoryear{Levin}{2007}]{Levin2007}
    Levin Y., 2007, MNRAS, 377, 159

%\bibitem[\protect\citeauthoryear{McDermott et al.}{1988}]{McDermott1988}
%   McDermott P.N., Van Horn H.M., Hansen C.J., 1988, ApJ, 325, 725

\bibitem[\protect\citeauthoryear{Messios et al.}{2001}]{Messios2001}
    Messios N., Papadopoulos D.B., Stergioulas N., 2001, MNRAS, 328, 1161

%\bibitem[\protect\citeauthoryear{Negele \& Vautherin}{1973}]{NV1973}
%   Negele J.W., Vautherin D., 1973, Nucl. Phys., A207, 298

%\bibitem[\protect\citeauthoryear{Piro}{2005}]{Piro2005}
%    Piro A.L., 2005, ApJ, 634, L153

%\bibitem[\protect\citeauthoryear{Poedts \& Kerner}{1991}]{PK1991}
%    Poedts S., Kerner W., 1991, Phys. Rev. Lett., 66, 2871

%\bibitem[\protect\citeauthoryear{Rincon \& Rieutord}{2003}]{RR2003}
%    Rincon F., Rieutord M., 2003, A\&A, 398, 663

%\bibitem[\protect\citeauthoryear{Reese et al.}{2004}]{RRR2004}
%    Reese R., Rincon F., Rieutord M., 2004, A\&A, 427, 279

\bibitem[\protect\citeauthoryear{Samuelsson \& Andersson}{2007}]{SA2007}
    Samuelsson L., Andersson N., 2007, MNRAS, 374, 256

%\bibitem[\protect\citeauthoryear{Schumaker \& Thorne}{1983}]{Schumaker1983}
%    Schumaker B.L., Thorne K.S., 1983, MNRAS, 203, 457

\bibitem[\protect\citeauthoryear{Sotani et al.}{2007a}]{Sotani2007a}
    Sotani H., Kokkotas K.D., Stergioulas N., 2007a, MNRAS, 375, 261 (Paper I)

\bibitem[\protect\citeauthoryear{Sotani et al.}{2007b}]{Sotani2007b}
    Sotani H., Kokkotas K.D., Stergioulas N., Vavoulidis M., 2007b, preprint (astro-ph/0611666) (Paper II)

\bibitem[\protect\citeauthoryear{Sotani et al.}{2007c}]{Sotani2007c}
    Sotani H., Colaiuda A., Kokkotas K.D., 2007c, preprint (gr-qc/0711.1518)

\bibitem[\protect\citeauthoryear{Sotani et al.}{2007d}]{Sotani2007d}
    Sotani H., Kokkotas K.D., Stergioulas N., 2007d, in preparation
    
%\bibitem[\protect\citeauthoryear{Strohmayer}{1991}]{Strohmayer1991a}
%   Strohmayer T.E., 1991, ApJ, 372, 591

%\bibitem[\protect\citeauthoryear{Strohmayer \& Watts}{2005}]{SW2005}
%   Strohmayer T.E., Watts A.L., 2005, ApJ, 632, L111

%\bibitem[\protect\citeauthoryear{Strohmayer \& Watts}{2006}]{SW2006}
%   Strohmayer T.E., Watts A.L., 2006, ApJ, 653, 593

   
\bibitem[\protect\citeauthoryear{Stergioulas et al.}{2004}]{Stergioulas2004}
    Stergioulas N., Apostolatos T.A., Font J.A., 2004, MNRAS, 352, 1089
    
\bibitem[\protect\citeauthoryear{Vavoulidis et al.}{2007}]{Vavoulidis2007}
    Vavoulidis M., Stavridis A., Kokkotas K.D., Beyer H., 2007, MNRAS, 377, 1553

\bibitem[\protect\citeauthoryear{Watts \& Strohmayer}{2006}]{WS2006}
   Watts A.L., Strohmayer T.E., 2006,  Advances in Space Research, in press
   (astro-ph/0612252)
   
%\bibitem[\protect\citeauthoryear{Watts \& Strohmayer}{2006}]{WS2006}
%   Watts A.L., Strohmayer T.E., 2006, ApJ, 637, L117




\end{thebibliography}
\end{document}